\documentclass[11pt]{article}
\usepackage{setspace, graphics, feynmp}
\newcommand{\be}{\begin{equation}}
\newcommand{\ee}{\end{equation}}
\newcommand{\bea}{\begin{eqnarray}}
\newcommand{\eea}{\end{eqnarray}}

\begin{document}
\begin{titlepage}
\def\thepage {}        

\title{Upper Limit on the Higgs Particle Mass}

\author{
Marko B. Popovic\thanks{Present address: 
Department of Physics, Lyman Lab, room 428, Harvard University, Cambridge, MA 02138. E-mail: marko@lorentz.harvard.edu}\\
Department of Physics, Boston University \\
590 Commonwealth Ave., Boston MA  02215}

\date{\today}

\maketitle

\bigskip
\begin{picture}(0,0)(0,0)
\put(295,250){BUHEP-01-15}
\put(295,235){hep-ph/0106355}
\end{picture}
\vspace{24pt}

\begin{abstract}

It is shown that the effective Standard Model theory with the physical Higgs mass lighter than $\sim 200$ GeV takes the form of an unbroken electroweak theory already at moderately high, $O(1$ TeV$)$, energy scales. No such transitional scale exists for the heavier Higgses. This result is independently obtained by two standard regularization methods.

\pagestyle{empty}
\end{abstract}
\end{titlepage}


\section{Introduction and synopsis}
\setcounter{equation}{0}

If a complete theory of Nature existed down to the smallest space-time distances, such a theoretical framework would most likely be free of infinities.
However, no such theory exists today. It is therefore not surprising that the ultraviolet divergences arise when quantum corrections with the highest momenta are considered. Hence, infinities encountered in the current theoretical framework are just a sign of lurking new physics embedded in the continuum. The current theoretical framework breaks down at a certain high-energy scale above which a new physics is required.\footnote{On dimensional grounds we believe that something new should happen at least around the Planck scale ($\approx 10^{19}$ GeV).} 

In practice, infinities are ``swept under the rug" as they should be. Dimensional regularization (or Euclidean hard cut-off method) provides a useful recipe for subtracting two infinities to obtain a finite, physically related quantity. In the case of renormalizable theory, infinities coming from every order of perturbation may be buried in the parameters of the same set of operators that one started with at the tree level. The Standard Model (SM) is one example of such a consistent theoretical structure.  

As a main instrument in the investigating endeavor, one may define the effective theory at some energy scale by the running effective Lagrangian which most suitably represents, at the classical tree level, physics at that particular energy scale. To examine physics at small distances, in the way similar to the Wilson's approach \cite{Wilson}, one may start with a known effective Lagrangian at the lowest energies and run it up to the highest energies. In this picture, SM renormalizability means that the running flow can be refined to any order of the perturbation theory without requiring new types of operators. It also means that the introduced running scheme should be, in principle, self-consistent down to the smallest attainable space-time distances at which point either the unacceptable vacuum arises (stability) \cite{triviality1, stability1, stability15, stability2, stability3, stability4}, the perturbation approach fails (perturbativity) \cite{perturbativity1, perturbativity2}, and/or some new physics takes a role. 

In this study it is investigated under which conditions the effective SM structure, when advanced toward the continuum, takes the form of an unbroken electroweak gauge theory. In other words, it is analyzed the possibility that there exists an energy scale at which the SM effective Higgs mass squared changes its sign. There is no definitive theoretical reason why this zero-temperature assumption has to be satisfied. However, the rationale for its appropriateness seems well supported. Clearly, the good high-energy behavior of the theory is guaranteed; in the ``unbroken" regime the Goldstone boson equivalence theorem is exactly right and the scalar propagator perfectly maintains unitarity. Moreover, for positive scalar quartic coupling this assumption guarantees that no other electroweak minimum can form at high energies. Finally, the fact that the loop correction is equal to the tree level value may suggest that theory is not tremendously finely tuned at the corresponding energy.\footnote{However, one needs to be careful when interpreting this possibility in respect to the fine-tuning problem. That will be addressed in detail in the subsequent paper.}    
 
In parallel, two independent techniques were utilized: $\overline{MS}$ scheme \cite{msbar}, applied to the effective potential \cite{effpotms1} analysis \cite{effpotms2, effpotms3}, and Euclidean hard cut-off scheme, applied to the generalized original Veltman's approach \cite{Higgs1}.

Using a notation adopted here, the tree level potential of neutral component of Higgs scalar doublet is 
\be
V(\phi)=-{m_H^2 \over 4} \phi^2 + {\lambda \over 8} \phi^4  \;\;\; , \label{jedan}
\ee
where $m_H^2=V^{(2)}|_{\langle\phi\rangle}$ is the tree level Higgs mass squared and $\langle\phi\rangle =v_{EW}=246.2$ GeV.\footnote{Alternatively, one may write $V(\psi)=-{m_H^2 \over 2} \psi^2 + {\lambda \over 2} \psi^4$ where $\langle\psi\rangle=v_{EW}/\sqrt{2}$ but then $m_H^2={1 \over 2} V^{(2)}|_{\langle\psi\rangle}$.} Hence, the running effective potential is defined as 
\be
V(\phi_R)=-{m_H^2(\Lambda\sim\phi_R) \over 4} \phi_R^2 + {\lambda(\Lambda\sim\phi_R) \over 8} \phi_R^4 \;\;\; , \label{dva}
\ee
with running effective parameters $m_H^2$ and $\lambda$. The connection with 
effective action at zero external momentum (i.e. the effective potential $V_{eff}$) is then simply
\be
V_{eff}(\phi_{cl})=V(\phi_R) \;\;\; , \label{tri}
\ee
where $\phi_{cl}$ is the classical field (on which the generating functional of 1-Particle-Ireducible Green functions depends) corresponding to the running field $\phi_R$. Obviously, it is the zero-temperature effective potential $V_{eff}$, and not some particular values of the running effective parameters $m_H^2$ and $\lambda$, that defines the vacuum structure of the theory. If the minimum of $V_{eff}$ is away from zero, the electroweak symmetry is broken. 

The original Veltman's approach \cite{Higgs1} can be used to describe the running effective potential in Eq. (\ref{dva}). Veltman reached the conclusion that by redefining mass terms and fields (i.e. running), the SM Lagrangian with one-loop corrections (and with only quadratic divergences considered) may be brought in a gauge invariant fashion to the same form as the tree level Lagrangian.\footnote{However, the sign of the running was not properly understood, which lead to wrong conclusion that the case with a dominant heavy top quark, as we today know is the actual case, should be considered unphysical.} This result was confirmed by Osland and Wu \cite{Higgs15} and refined with additional logarithmic terms at the one-loop level by Ma \cite{Higgs2} in $R_{\xi}$ gauge. Moreover, the running of all the couplings of interest was included \cite{Higgs3}. In addition, the higher-loop contributions to quadratic running were calculated in recursive manner \cite{Higgs4, Higgs5}. The lesson of this conceptual view is that SM suffers from quadratic divergences and that the running effective Higgs mass squared at one-loop level satisfies \cite{Higgs1, Higgs2}
\bea
{d m_H^2 \over d\Lambda^2} = {3 g_2^2 \over 64 \pi^2 M_W^2}
 \left[ m_H^2 + 2 M_W^2 + M_Z^2 - 4\sum_f \left( {n_f \over 3} \right)
m_f^2 \right] + \nonumber \\+{3 g_2^2 \over 64 \pi^2 M_W^2} {m_H^2 \over {2 \Lambda^2}}\left[m_H^2 - 2 M_W^2 - M_Z^2 +
2\sum_f \left( {n_f \over 3} \right) m_f^2 \right] \;\;\; , \label{cetiri}
\eea 
where $n_f=3 \, (1)$ for quarks (leptons). Using the tree level SM relations, the above mass formula may be rewritten in terms of the gauge couplings $g_i$'s, Yukawa couplings $g_f$'s, and quartic coupling $\lambda$ as
\be
{d m_H^2 \over d\Lambda^2} = -{1 \over 32 \pi^2} \left[ 12 g_t^2- 6 \lambda-{9 \over 2} g_2^2 - {3 \over 2} g_1^2 \right] + \ldots \;\;\; . \label{dodato}
\ee

Conversely, in the $\overline{MS}$ scheme, the scalar mass squared, $m^2(t)$, is running just logarithmically! Does this mean that SM in the $\overline{MS}$ scheme does not suffer from quadratic divergences embodied in the running effective Higgs mass squared? The answer is no, and to show this, it is useful to consider the effective potential $V_{eff}$.   

\begin{figure}
\resizebox{13cm}{9cm}{\includegraphics{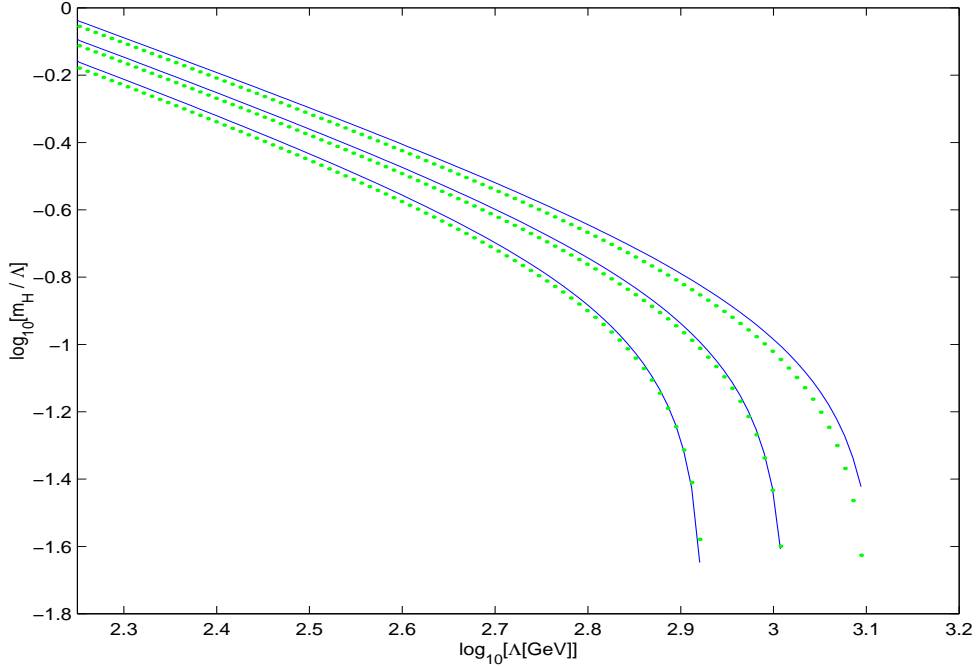}}
\caption[higgs0]{\small The running effective Higgs masses obtained from the two methods, $\overline{MS}$ scheme (solid lines) and hard cut-off scheme (dotted lines), for a few representative physical Higgs masses: $120$, $140$, and $160$ GeV.}
\label{adriatic0}
\end{figure}

Following the approach of \cite{stability2, stability3} in the $\overline{MS}$ scheme and in the 't Hooft-Landau gauge, the renormalization group improved one-loop effective potential \cite{effpotms2, effpotms3} is
\bea
V_{eff}=V_0+V_1 \;\;\; \mbox{where} \;\;\; V_0=-{1 \over 2} m^2(t)\phi^2_R(t)+{1 \over 8}\lambda(t)\phi^4_R(t) \;\;\; \label{cetiriipo} \\ \mbox{and} \;\;\; V_1=\sum_{i=1}^5 {n_i \over {64 \pi^2}}(\kappa_i \phi^2_R(t) - \kappa_i^{'})^2 \left[log{{\kappa_i \phi^2_R(t) - \kappa_i^{'}} \over \mu^2(t)}-c_i \right] + \Omega(t) \;\;\; . \label{pet}
\eea
The values of the parameters $n$, $\kappa$, $\kappa'$, and $c$ are given in reference \cite{stability2, stability3}. Contribution to the cosmological constant is denoted by $\Omega$ and is assumed irrelevant for the current calculations. Classical and running Higgs fields are related as
\be
\phi_R(t)=exp\left[-\int_0^t \gamma(t')dt'\right]\phi_{cl} \; , \;\; \mbox{where}  \;\;\; \gamma(t')={3 \over {16 \pi^2}}\left(g_t^2(t')-{1 \over 4}g_1^2(t')-{3 \over 4}g_2^2(t')\right) \label{sest}
\ee
is the anomalous dimension, and $g_t$ is the top Yukawa coupling. Using Eqs. (\ref{dva}-\ref{tri}) the running effective Higgs mass squared is extracted from Eqs. (\ref{cetiriipo}-\ref{sest}) as
\be
m_H^2=-{{4V_{eff}(\phi_{cl})-{\lambda \over 2} \phi_R^4} \over \phi_R^2} \label{sedam}
\ee

This is a quantity that should be compared with the quadratically unstable Higgs mass squared, Eq. (\ref{cetiri}), in the method developed from the original 
Veltman's approach! In Fig. 1 the running effective Higgs mass squared obtained from the two methods are compared for a few representative physical Higgs masses.

After defining the tools, we proceed to their application.
 
The well established limits on the physical Higgs mass are obtained from the requirement that the electroweak low energy vacuum is perserved (stability bounds) \cite{triviality1, stability1, stability2, stability3, stability4} and that the perturbation theory is valid (perturbativity bounds) \cite{perturbativity1, perturbativity2} up to the Planck scale. The stability yields a lower bound on the Higgs mass (roughly in between $130$ and $140$ GeV \cite{stability2, stability3, stability4}), while the perturbativity gives an upper bound (roughly in between $170$ and $180$ GeV \cite{perturbativity1, perturbativity2}). However, these are the bounds from the desert-like scenarios. What if there is some new physics below the Planck scale? The above window on Higgs mass is then simply inadequate. In that case, the perturbativity and stability curves, as functions of the physical Higgs mass, trace only the highest possible energy scale at which some new physics must enter the description. Therefore, the physical Higgs mass is left largely unconstrained up to the triviality bound \cite{triviality1, triviality2, triviality3} of roughly $O(0.5$TeV$)$, when the heavy Higgs mass becomes approximately equal to the scale of the one-loop scalar quartic coupling's Landau pole. Is there any other criteria (unrelated to the desert-like scenario) that may be applied? 

The assumption contrived here is that parameter $m_H^2(\Lambda)$ needs to become zero at some energy scale,\footnote{While assuming that running according to SM gives an appropriate description up to the same energy scale $\Lambda$.} from now on called the Higgs mass zero crossing (HMZC) scale. Although there is no strict theoretical necessity for this assumption to be unavoidably satisfied nonetheless, intuitively, it seems to be very natural.\footnote{And many new physics models need to satisfy this assumption. An example is the minimal top-mode SM \cite{lepirad}.}  Basically this assumption represents an analogue of the Coleman-Weinberg conjecture \cite{effpotms1} where the bare scalar mass ($m^2_{bare}$ however, and not $m_H^2(\Lambda)$ as here) is zero and electroweak symmetry breaking is essentially governed by the quantum corrections.\footnote{In the framework of a renormalizable theory, this intriguing conjecture unfortunately fails \cite{hamshowed} due to the large, experimentally measured top quark mass; not due to the originally thought large scale at which the one-loop approximation becomes unreliable.} 

The approaches and ideas somewhat similar\footnote{Though, in many aspects completely different.} to those presented here were addressed already in the past \cite{hamshowed, cvetic}. However, the upper bound on the Higgs mass as obtained there \cite{hamshowed, cvetic} is generally $50-100$ GeV larger than the upper bound determined in this study. 

The running Higgs masses squared as obtained from the $\overline{MS}$ effective potential analysis \cite{effpotms2, effpotms3} as well as from the Euclidean hard cut-off generalized Veltman's approach are scrutinized on the similar footing: at the one-loop level with logarithmic terms included and with running of all the couplings of interest at the two-loop level, i.e. in the next-to-leading-log (NTLL) level approximation. In the next section, the upper bound of 
the physical Higgs mass is established from the requirement that the HMZC scale exists. 

\section{Results}

The strong coupling and the top pole mass are taken to be $\alpha_s (M_Z)=0.1182$ and $m_t=175$ GeV respectively. The matching condition for the running top Yukawa coupling is identical to the one in \cite{stability2, stability3}.
The one-loop level matching conditions for $m^2$ and $\lambda$ in the $\overline{MS}$ effective potential approach, and $m_H^2$ and $\lambda$ in the hard cut-off generalized Veltman's approach were obtained from the standard requirement that
\be
V_{eff}^{(1)}|_{\langle\phi\rangle}=0 \;\;\; \mbox{and} \;\;\; V_{eff}^{(2)}|_{\langle\phi\rangle}=m_H^2(0)=m_H^2(pole)-\Delta\Pi(m_H^2(pole))
\ee 
where $\Delta\Pi(m_H^2(pole))=Re\left[\Pi(m_H^2(pole))-\Pi(0)\right]$ with $\Pi$ 
being the renormalized self-energy of the Higgs boson. The reader is directed to reference \cite{stability2} for more details. That approach has been closely followed here with the main results reconfirmed with a very good precision. In the case of Euclidean hard cut-off generalized Veltman's approach the main difference in the matching procedure results from the different form of the running effective potential Eq. (\ref{dva}). The running of the Higgs mass squared is given by Eq. (\ref{cetiri}-\ref{dodato}). Although lengthy, the matching procedure in the generalized Veltman's approach is rather trivial - we restrain ourselves from going to any details here. 

After the matching conditions are properly set, the gauge and Yukawa (top is only relevant) couplings are let to run at the two-loop level \cite{2lrunning, effpotms2}. In Fig. 2, the stability, perturbativity, and HMZC scales (those with $dm_H^2/d\Lambda^2|_{\Lambda_{HMZC}}<0$), as obtained here, are shown for both methods. The stability is justified by the presence of an unacceptable, deeper minimum of the effective potential at the high 
\begin{figure}
\resizebox{13cm}{9cm}{\includegraphics{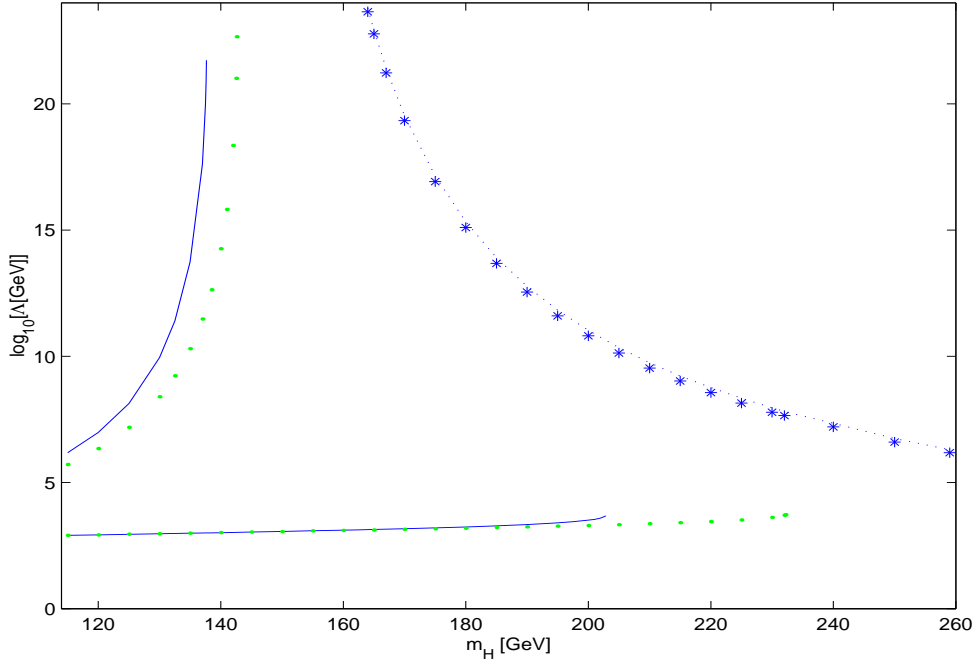}}
\caption[higgs1] {\small The $\overline{MS}$ scheme (solid line) and hard cut-off method (dotted line) stability curves are shown in the upper left corner. The perturbativity curve in the $\overline{MS}$ scheme (stars) together with $\overline{MS}$ scheme-specific high-energy HMZC scales (dotted line) is shown in the upper right segment of the figure. The regular HMZC scales in the $\overline{MS}$ scheme (solid line) and hard cut-off method (dotted line) are presented in the lower part of the figure.}
\label{adriatic1}
\end{figure}
energies. This curve traces the energy scale at which the sharp transition (energy at which roughly $V_{eff}^{(2)}\rightarrow 0$) occurs as a function of 
the physical Higgs mass. The perturbativity is defined via $\lambda=4$ limit\footnote{Roughly $1/6$ of the two-loop level quasi-fixed point $\lambda_{FP}\approx 24$.} \cite{lambda1}, i.e. the perturbativity curve traces that value of the running quartic coupling as a function of the physical Higgs mass. As shown in \cite{lambda1} the convergence of a perturbative series of high-energy Higgs and Goldstone-boson scattering amplitudes is lost when $\lambda(\Lambda)/2$ is of order  $2 - 2.3$ and larger. Under the assumption that the Planck scale defines the SM validity, and under the stability and perturbativity constraints, the window on the allowed Higgs masses, as obtained here in the $\overline{MS}$ scheme, is $137 \leq m_H^{physical} \;^{+}_{-} 5  \leq 171$ GeV.\footnote{Roughly $136 \leq m_H^{physical} \;^{+}_{-} 5  \leq 178$ GeV if Planck scale is traded for the GUT scale $\sim 10^{16}$ GeV.} This result is in a very good agreement with the previous findings \cite{stability2, stability3, perturbativity1, perturbativity2}.\footnote{Somewhat smaller upper bound from the perturbativity considerations (with $\lambda=4$ delimiting value) in \cite{stability4} is caused likely by larger $\alpha_s=0.123$ and slightly different top Yukawa matching. However, the results almost overlap at one $\sigma$ level.}  

In Figure 2, in addition to the regular, low energetic (roughly $O(1$TeV$)$) HMZC scales, the $\overline{MS}$ approach-specific high-energy HMZC scales have also been shown (the HMZC scales with $dm_H^2/d\Lambda^2|_{\Lambda_{HMZC}}>0$ are omitted). However, these high-energy HMZC scales (with $dm_H^2/d\Lambda^2|_{\Lambda_{HMZC}}<0$) always exceed the perturbativity limiting scales and therefore are considered unphysical. Even if the perturbativity bound is pushed to somewhat larger values of quartic coupling (for example $\lambda=6.4$ as in \cite{luwei} or $\lambda=8$ as in \cite{lambda2}), this transition should not be trusted due to the non-decoupling nature of the $\overline{MS}$ scheme and the closeness in the energy scales.

The HMZC scales in both methods are presented in Fig. 3. While there is an excellent agreement for the lighter Higgses (roughly below $190$ GeV), tangible difference arise for the heavier scalars.\footnote{It will be very interesting to see how might the addition of the two-loop level contributions change this results. In general, the presence of the heavy Higgs should stimulate more refined analysis for both methods because, for example, the $\lambda^2$ terms become non-negligible.} A requirement that a HMZC scale exists, yields an upper bound of $203^{+14}_{-3}(232^{+17}_{-8})$ GeV  on the 
physical Higgs mass in the $\overline{MS}$ scheme (Euclidean hard cut-off method). However, the range of the transitional HMZC scales, $\Lambda_0$, is bound with an exceptional agreement in both methods to the window $10^{2.9} \leq \Lambda_0 < 10^{3.7}$ GeV as a function of the physical Higgs mass.

The variations in the physical Higgs mass upper bound due to the variation in $\alpha_s$ and $m_t$ in the linear approximation are
\be
\delta m_H [\mbox{GeV}]=1.40\delta m_t[\mbox{GeV}]-360\delta\alpha_s \;\;\; \mbox{and} \;\;\; \delta m_H [\mbox{GeV}]=1.80\delta m_t[\mbox{GeV}]-70\delta\alpha_s 
\ee
in the $\overline{MS}$ scheme and generalized Veltman's method, respectively.

The errors are determined in a rather conservative manner. They are obtained from the separate variations of the matching conditions, in the range from the one-loop level matching to the tree level matching, for the three main parameters: $m_H^2$ (or $m^2$), $\lambda$, and $g_t$. Top quark Yukawa coupling matching condition is mainly responsible for the upper errors. The quartic coupling matching condition tends to bring the results of the two methods closer. The results are rather insensitive to the variations in the mass squared matching conditions. Following the same logic as in \cite{stability2, stability3} for the $\overline{MS}$ scheme, the $O(3$ GeV$)$ uncertainty is also incorporated in response to the requirement that the effective potential $V_{eff}$ must be renormalization scale independent. Finally, the separate errors on the physical Higgs mass have been added in quadrature. At list in numerical sense one could also calculate the joint limit for the two independent methods. The joint limit is $225^{+6 \; (11)}_{-7 \; (12)}$ GeV at $69\%$ ($90\%$) c.l.. 

Another way to estimate the error in $\overline{MS}$ scheme is to again reproduce the effective potential analysis but with the new set of matching conditions, as provided by the reference \cite{perturbativity2}. These HZMC scales are also presented in Fig. 3 for the heavier Higgses. The central value is now depicted as the mean of the old and new result while the error, in the Gaussian approximation, is rounded off roughly as a difference between the two. The upper bound on the physical Higgs mass is now tighten to $205^{+}_{-}4$ GeV.

\begin{figure}
\resizebox{13cm}{9cm}{\includegraphics{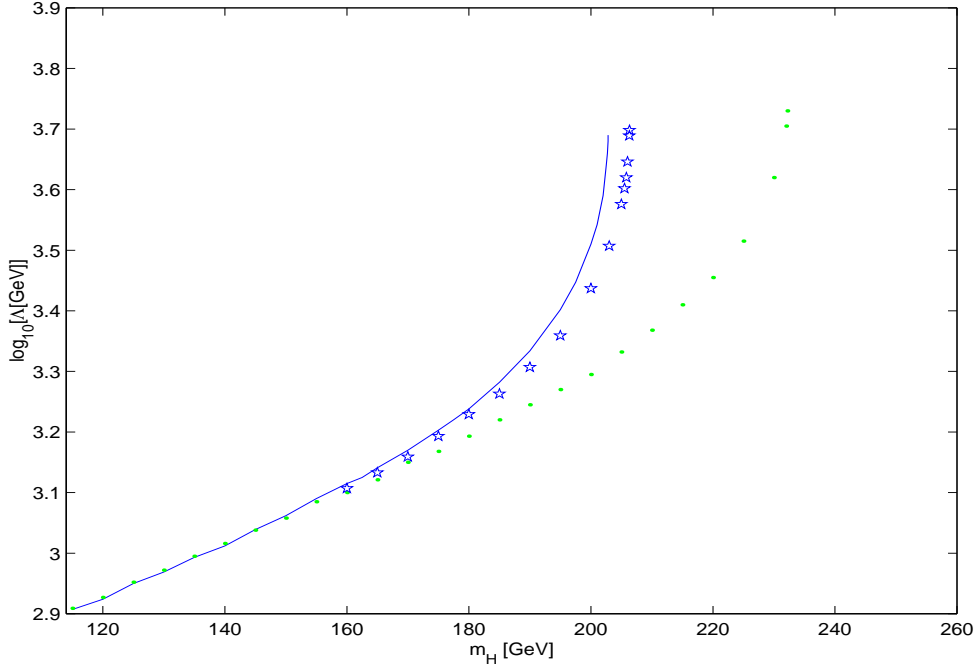}}
\caption[higgs2] {\small HMZC scales ($m_H^2(\Lambda_{HMZC})=0$) in the $\overline{MS}$ scheme (solid line) and Euclidean hard cut-off method (dotted line). Heavy Higgs portion of the HMZC scales with matching condition as in \cite{perturbativity2} are presented by pentagrams. }
\label{adriatic2}
\end{figure}
The results obtained in this work lead to some consequences on the Minimal Supersymmetric Standard Model (MSSM), too. While the more sensible MSSM may be distinguished via the MSSM decoupling scales (defined here by the stop mass, $m_T$) which are below the (SM) HMZC scales, the less sensible MSSM may be characterized by the MSSM decoupling scales which are above the HMZC scales. Using the approximate relation $m_H^2\leq m_Z^2+{3G_F \over {\sqrt{2}\pi^2}}m_t^4\ln{m_T^2 \over m_t^2}$ (see for example \cite{awson}) for the radiatively corrected Higgs mass in the MSSM, delimiting scale is found to be approximately $900$ GeV (in agreement with the standard $1$ TeV scale as obtained from the naturality considerations \cite{barbi}) and corresponding to $m_H\approx127$ GeV (in agreement with findings in \cite{findings}). In the less natural case ($m_T\geq1$ TeV), the MSSM and SM frameworks share the same results on the HMZC scales. However, the gap between the HMZC scales and MSSM decoupling scales grows rapidly with heavy Higgs masses. The MSSM decoupling scale is roughly ten times larger than the HMZC scale for $m_H \sim 170$ GeV and around hundred times larger at the verge of the Higgs masses for which the HMZC scale exists.\footnote{However, more sophisticated result $ m_H^2 \leq M_Z^2 + { 3 g_2^2 M_Z^4 \over 16 \pi^2 M_W^2 } \left[ \left( {{2 m_t^4 - m_t^2 M_Z^2} \over M_Z^4} \right) \ln{m_T^2 \over m_t^2} + {m_t^2 \over 3 M_Z^2} \right]$ for the radiatively corrected Higgs mass in the MSSM \cite{awson} yields even tighter delimiting scale, $\approx 860$ GeV, corresponding to the $m_H\approx122$ GeV (and already for $m_H \sim 160$ GeV the MSSM decoupling scale becomes roughly ten times larger than the HMZC scale).}

\section{Conclusion}

The upper mass limit on the SM Higgs is predicted from the requirement that the running effective Lagrangian, at some high energy, takes the form of an essentially unbroken theory of electroweak interactions. The result is obtained, and the main conclusions independently confirmed, using two different regularization methods: $\overline{MS}$ scheme (applied to the effective potential analysis) and Euclidean hard cut-off scheme (applied to the generalized Veltman's approach). 

Higgs is found to be lighter than $203^{+14}_{-3}(232^{+17}_{-8})$ GeV  in the $\overline{MS}$ scheme (Euclidean hard cut-off method). Therefore, the mass window on the SM Higgs, obtained previously from the considerations of the desert-like scenarios, stability and perturbativity constraints, is now broadened through the less restrictive assumption and without reference to the Planck or some other in advanced fixed, high energy scale. The upper mass limit determined here is much tighter than the one, $O(0.5$ TeV$)$, that arise from the triviality considerations. 

The transitional HMZC scale is found to emerge in the window $10^{2.9} \leq \Lambda_0 \leq 10^{3.7}$ GeV as a function of the physical Higgs mass. Interestingly, this shows good agreement with the familiar result, $\Lambda_0 = 4 \pi \kappa v_{EW} / \sqrt{2} (\sim 10^{3.3} GeV \; \mbox{for} \; \kappa=1$) derived from the dimensional analysis \cite{nlsigmam} applied to the non-linear sigma model and strongly interacting chiral Lagrangians.

If the mass of the soon-to-be discovered SM Higgs particle happens to agree with current limits \cite{ewpmHiggs} from the electroweak precision data, then there is a good chance that Nature is indeed characterized by an $O(1$ TeV$)$ transitional scale. This frontier scale will then be the next step in our quest for the ultimate theory of Nature.

\section*{ Acknowledgments }
\indent

The author would like to thank his spouse Jelena Popovic for all her love and
support during a time when this work was in progress. The author is also
expressing his gratitude to R. S. Chivukula, G. Cvetic, H. Herr,
C. Hoelbling, K. Lynch, D. Marfatia, M. Schmaltz, E. H. Simmons and T. Rador on all their questions and answers, disagreements and agreements. This work was supported in part by the National Science Foundation under grant PHY-9501249, and by the Department of Energy under grant DE-FG02-91ER40676.

\end{document}